\documentstyle[11pt,moriond,epsfig]{article}

\def\be{\begin{equation}}
\def\ee{\end{equation}}
\def\bea{\begin{eqnarray}}
\def\eea{\end{eqnarray}}

\begin{document}
\vspace*{4cm}
\title{A review of the CMS discovery potential for MSSM SUSY \\massive Higgses A and H}

\author{A.~Van Lysebetten}

\address{DAPNIA/SPP, Centre d'\'{e}tudes de Saclay,\\91191 Gif sur Yvette CEDEX,
France}

\maketitle\abstracts{The present understanding of the expected MSSM SUSY heavy neutral $A^0$ and $H^0$ Higgs reach at CMS is reviewed. Two Higgs decay channels are considered in this talk: $A,H$ $\to$ $\tau \tau$ and $A,H$ $\to$ $\chi_2^0 \chi_2^0$. Under the assumption that neutralinos and sleptons are not too heavy, the channel $A,H$ $\to$ $\chi_2^0 \chi_2^0$ provides an interesting and complementary channel to the $\tau \tau$ channel. The expected reach in the MSSM parameter space for both channels as well as their discovery potential at the CMS detector will be presented.
}
\section{Introduction}
In the Minimal Supersymmetric Standard Model (MSSM) two Higgs SU(2) doublets are assumed to produce a spontaneous breakdown of SU(2) $\times$ U(1), which leaves five physical degrees of freedom: two neutral CP even Higgs bosons $h$ and $H$, one neutral CP odd $A$ and two charged Higgs bosons $H^{\pm}$. \\
The most promising channel to study the massive neutral Higgs bosons A and H at the Large Hadron Collider (LHC) seem to be $A,H \to \tau\tau$. This channel has been studied very intensively~\cite{cmshh,cmshl,cmsll1,cmsll2} and shows good perspectives for large and intermediate $\tan\beta$ regions. This channel has also the advantage of mass reconstruction with a relative good resolution. The analysis techniques for the different $\tau\tau$ decay channels will be reviewed in Section~\ref{sec:tautau}.      \\
In the previous channel it is assumed that sparticles are heavy (of the order of 1 TeV), and therefor do not participate in the decay processes. If we assume that neutralinos and/or charginos are light, decay modes to these particles opens. In Section~\ref{sec:neutra} we will investigate the possibility to observe $A,H \to \chi_2^0\chi_2^0$. We will demonstrate that it allows access to a region in the MSSM parameter space complementary to the region covered by the channel $A,H \to \tau \tau$.  
\section{Higgs production and decay at the LHC}
In all the studies presented below the generator {\tt Spythia}~\cite{Spythia} was used for the generation of the signal events and the background events were generated with the {\tt Pythia}~\cite{Pythia} program. The detector response of the CMS detector was simulated with {\tt Cmsjet}~\cite{cmsjet}, the fast simulation package, based on a detailed {\tt Geant} simulation including Level 1 and Level 2 trigger effects. 
\section{The $\tau \tau$ channel}
\label{sec:tautau}
The $\tau$ particle being short lived and having hadronic and leptonic decay modes, it gives rise to three different final states: {\it 2$\tau$ jets}, {\it$\tau$ jet + lepton} and {\it2 leptons}. The analysis techniques for the two first final states will be discussed in the next sections (Sections~\ref{subsec:tauhadro} to~\ref{subsec:taulept}). For the hadronic $\tau$ decays only 1 prong decays have been included in the present analyses.
\subsection{Two $\tau$ particles decay hadronically: $A,H \to h^+ + h^- + X$}
\label{subsec:tauhadro}
\begin{figure}
\begin{center}
\psfig{figure=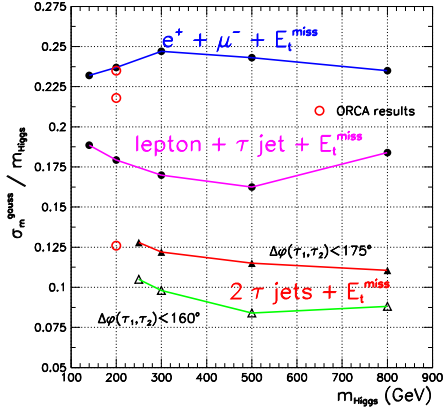,height=6.cm}
\end{center}
\caption{The Higgs mass resolution as function of the mass m$_A$ for the different $\tau\tau$ decay channels.
\label{fig:massresol}}
\end{figure}

\begin{figure}
\begin{centering}
\psfig{figure=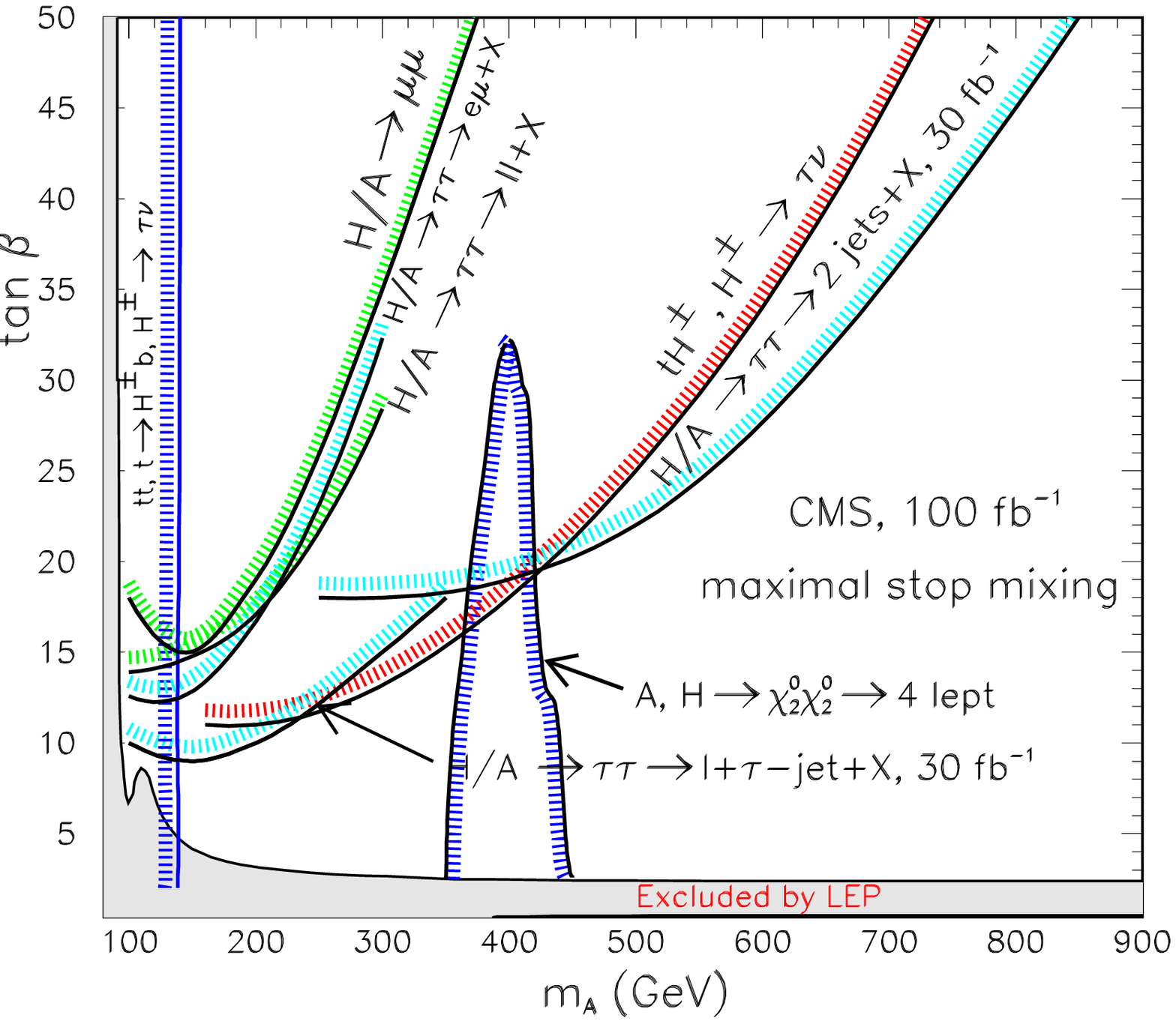,height=6cm,width=6cm}
\psfig{figure=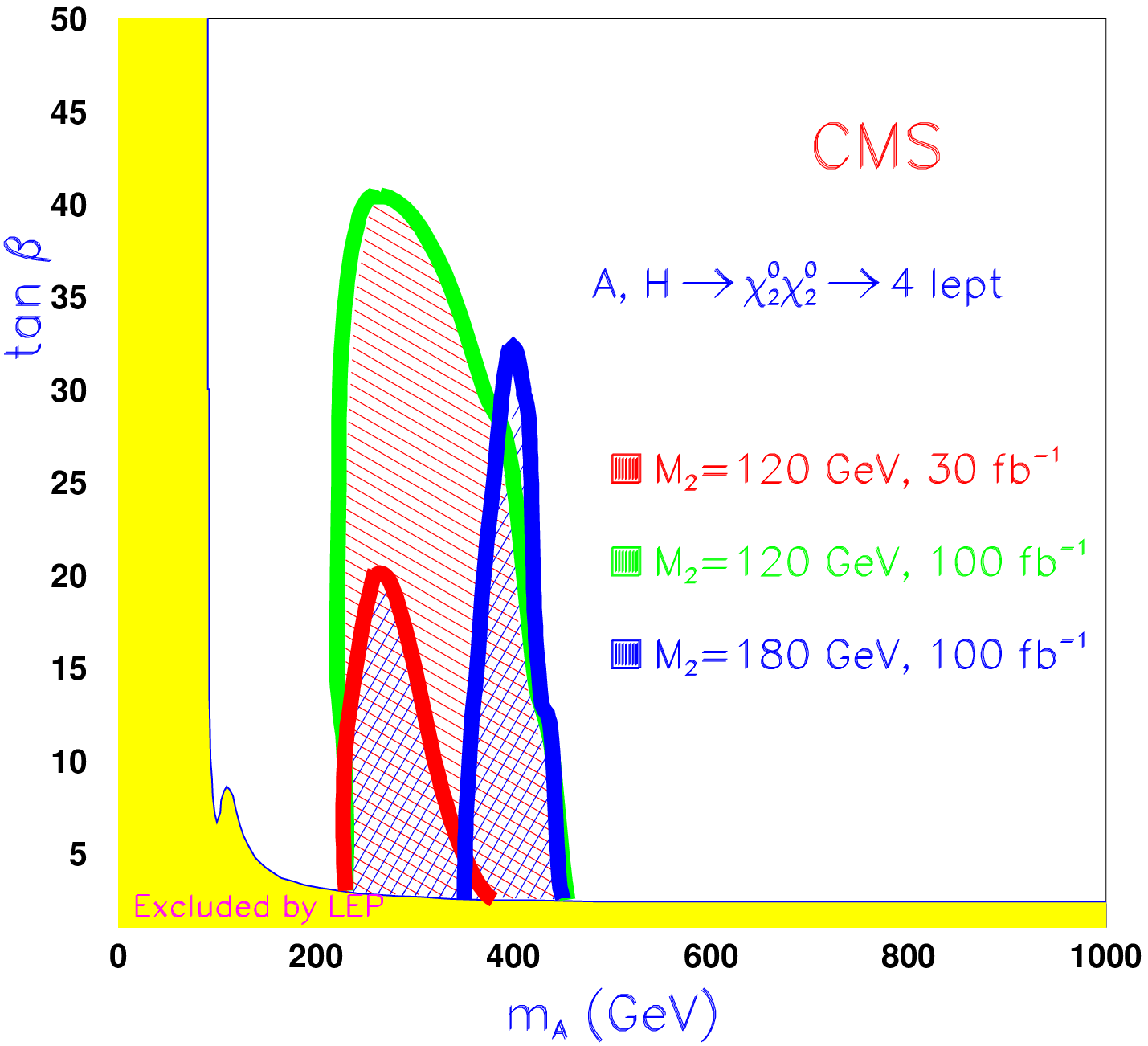,height=6cm,width=6cm}
\caption{(a,left)The 5$\sigma$ discovery potential for the different $\tau\tau$ channels and the $\chi\chi$ channel for the following parameters: M$_1$ = 90 GeV, M$_2$=180 GeV, $\mu$=-500 GeV, m$_{\tilde{l}}$=250 GeV and m$_{\tilde{q},\tilde{g}}$=1 TeV;(b,right)The 5$\sigma$ discovery potential for the $\chi\chi$ channel for the same parameters and different values of M$_2$ and M$_1$.
\label{fig:disc}}
\end{centering}
\end{figure}

The final state consist of two isolated hard hadrons (plus $\pi^0$'s). These events are hard to trigger on as they are overwhelmed by the QCD di-jet background. The irreducible backgrounds are the Drell-Yan process $Z/\gamma^{*} \to \tau\tau$ and $t\bar{t}$ production with $W\to\tau\nu$. \\
Selection criteria were developed according to the typical topology and designed in order to reduce as much of the background processes as possible. Events are required to have at least two calorimetric jets with $E_T>$ 60 GeV within $|\eta|<2.5$. These candidates are then defined to be $\tau$ jet if it contains one isolated hard hadron within $\Delta R <$ 0.1 from the calorimeter jet axis. A cut on $p_T^h>$ 40 GeV is chosen in order to reject the QCD background. The isolation criteria is based on a cone algorithm, requiring no other track with $p_T>$ 1 GeV than the hard track in a cone of $\Delta R<$ 0.4 around the calorimeter jet axis. The two hard tracks from $\tau^+$ and $\tau^-$ have an opposite sign, while no such strong charge correlation is expected for QCD events. Events are triggered with a specially developed two-jet trigger with full trigger efficiency for $E_T^{jet}>$ 60 GeV. Due to the large Higgs masses considered here the missing transverse energy $E_T^{miss}$ from the neutrinos is expected to be significant, although the two $\tau$'s tend to be in a back-to-back configuration resulting in a partial compensation. A cut $E_T^{miss}>$ 40 GeV reduces the QCD background,suppressing it well below the irreducible $Z/\gamma^* \to \tau\tau$ background. For the largest masses this cut can be increased to $E_T^{miss}>$ 60 GeV.\\
Despite the two escaping neutrinos the Higgs mass can be approximately reconstructed from the two $\tau$ jets measured in calorimetry and from the $E_T^{miss}$ by projecting the $E_T^{miss}$ vector on the directions of the reconstructed $\tau$ jets. In order to reconstruct the Higgs mass the angle in the transverse plane $\Delta \phi$ between the two $\tau$ jets is required to be smaller than 175 degrees. This is rather costly for the signal as the two jets in the signal events are predominantly in a back-to-back configuration. The mass resolution, obtained by a Gaussian fit, is shown in Figure~\ref{fig:massresol} as function of the Higgs mass. The mass resolution can be significantly improved by reducing the upper limit in $\Delta \phi$ as is shown in Figure~\ref{fig:massresol} for $\Delta \phi<$160 degrees. As can be seen from this figure the mass resolution is the best in this $h^+h^-$ channel and the worst in the leptonic channel. The mass resolution is dominated by the precision of the $E_T^{miss}$ measurement and the resolution is better in the channels with a smaller fraction of energy carried away by neutrinos. This figure includes the first results (circles) obtained by a detailed simulation study, they seem in good agreement with the fast simulation results. A final cut is made on the ability to reconstruct the Higgs mass and the demand that this reconstructed mass lies within a certain interval (for example when the generated Higgs mass m$_A$= 300 GeV the reconstructed mass should lie in the interval 230 GeV$< \rm{m}_{\tau\tau} <$ 350 GeV.). \\
Figure~\ref{fig:disc}(a) shows the 5 $\sigma$ significance discovery contours, with $\sigma \equiv \frac{S}{\sqrt{S+B}}$ where S is the number of signal events and B the number of selected background events, for SUSY Higgs $A$ and $H$ as a function of m$_A$ and $\tan\beta$ for an integrated luminosity of 30 fb$^{-1}$ (without pile-up) under the no-stop mixing scenario. 
\subsection{One $\tau$ particle decays hadronically, the other leptonically: $A,H \to l + h + X$}
\label{subsec:taulept}
In this case the $l + \tau jet + E_T^{miss}$ final state with one isolated lepton and an hadronic $\tau$ jet is studied. Only $\tau$ jets decaying into one charged hadron are considered. Large backgrounds arise from $Z/\gamma^* \to \tau\tau$, $W+jet$ events, $t\bar{t}$ events with real or faked $\tau$'s and from $b\bar{b}$ events. \\ 
Events are triggered with a lepton + jet trigger, with an acceptable trigger rate and efficiency for the thresholds chosen here. The selection criteria are the following: one isolated lepton, $p_T^l>$ 15 GeV, $|\eta|<$ 2.4; one isolated $\tau$ jet, $E_T >$ 40 GeV, $|\eta|<$ 2.4; 60 degrees $<\Delta\phi(lepton-\tau jet)<$ 175 degrees; $E_T^{miss}> $ 20 GeV; m$_T$(lepton,$E_T^{miss}$) $<$  30 GeV; and a window in the reconstructed Higgs mass depending on m$_A$. Lepton isolation is performed in the tracker requiring no other track with $p_T>$ 1 GeV in a cone of $\Delta R<$0.3 around the lepton. A strong lepton isolation reduces significantly the very large $b\bar{b}$ background. The upper limit cut on $\Delta\phi$ is necessary to reconstruct the Higgs mass with a good resolution. The lower cut reduces the $t\bar{t}$ rate and the W + jet background. The cut on the missing transverse energy is relatively costly for the signal but is still useful against $Z/\gamma^* \to \tau\tau$ background. Backgrounds with real $W$ bosons can be effectively reduced by an upper bound on the transverse mass m$_T(lepton,E_T^{miss})$ calculated from the lepton and the $E_T^{miss}$ vector. The Higgs mass can be reconstructed from the lepton, $\tau$ jet and the $E_T^{miss}$ vector (see Section~\ref{subsec:tauhadro}). An overview of the mass resolution for this channel can be found in Figure~\ref{fig:massresol}. At this stage of the selection procedure the mass peak is not clearly visible on top of the $Z/\gamma^* \to \tau\tau$ background.\\
Further reduction of the overwhelming $Z/\gamma^*$ background can be obtained using the associated production processes $b\bar{b}A$ or $b\bar{b}H$. Therefor a $b$ tagging procedure was developed. The $b$ jet candidates are jets with $E_T>$ 20 GeV within $|\eta|<$2.5. The jet is defined as a $b$ jet if there are at least two tracks with $p_T>$ 2 GeV and $S_{IP}>$3 inside the jet within $\Delta R <$0.6. The parameter $S_{IP}$ is the significance of the Impact Parameter (IP) in the transverse plane defined as IP over its error. An upper bound on the impact parameter of 1 mm is applied to reduce the $K_s^0$ and $\Lambda^0$ contamination. The events with a $b$ jet can also be tagged by the presence of muons from $B$ decays. The combined tagging efficiency for signal events is small ranging from 10 to 24 $\%$ depending on m$_A$ and $\tan\beta$. The $Z/\gamma^*$ background is now strongly reduced to the level of the signal with a tagging efficiency of 1.3$\%$. The $A \to \tau \tau$ peak now becomes clearly visible on top of the background. Figure~\ref{fig:disc}(a) shows the 5$\sigma$ significance discovery contour for the SUSY higgs going to $\tau\tau$ decaying into a jet and a lepton as a function of m$_A$ and $\tan\beta$ for 30 fb$^{-1}$. 
\section{The neutralino neutralino channel}
\label{sec:neutra}
In the study above it is assumed that the sparticles are too heavy to participate in the decay process of the SUSY Higgses. If light neutralinos ($\chi^0$), charginos ($\chi^\pm$) and sleptons ($\tilde{l}$) exist and have low masses the SUSY Higgss could decay into these particles. As free parameters the CP-odd Higgs mass m$_A$, the Higgs VEV ratio $\tan\beta$, the Higssino mass parameter $\mu$, the bino mass parameter M$_1$, the wino mass parameter M$_2$, the slepton mass m$_{\tilde{l}}$ and the squark and gluino masses m$_{\tilde{q},\tilde{g}}$ are taken. Light neutralinos and charginos are considered. The renormalisation group relation M$_2$ $\equiv$ 2M$_1$ is used. The large $|\mu|$ scenario (M$_1<$M$_2< |\mu|$) is favoured in models where $\chi^0_1$ is the dark matter candidate (mSUGRA). For large $|\mu|$ values $\chi_2^0$ is rather wino and $\chi^0_1$ is bino like. Therefore is holds that m$_{\chi^0_1}$ $\approx$ M$_1$ and m$_{\chi^0_2}$ $\approx$ M$_2$. Also sleptons are taken light. The masses of squarks and gluinos are kept at 1 TeV. \\
Supersymmetric decay modes of the heavy Higgses are studied: $A,H \to \chi\chi$. For m$_A<$500 GeV the probability to decay into $\chi^+_1\chi^-_1$ is the highest (about 20 \% of the total BR), however it produces a final state with only 2 leptons, and is thus overwhelmed by various backgrounds. The second best sparticle mode is $\chi^0_2\chi^0_2$. This channel can provide a 4 lepton final state.  The next-to-lightest neutralino $\chi^0_2$ decays into two fermions and the lightest neutralino. To obtain a clean signal we will focus on the case where the neutralino decays into two leptons: $\chi^0_2 \to l^+ l^-\chi^0_1$, with $l=e {\rm or} \mu$. \\
The parameter space was scanned and the most optimal production cross section was found for the following parameter set: M$_1$= 60 GeV, M$_2$=120 GeV, $\mu$ = -500 GeV and m$_{\tilde{l}}$=250 GeV. For the background processes {\tt Pythia} was used, while the signal events were generated with {\tt Spythia}. The following SM backgrounds give rise to real or fake leptons in the final state: $ZZ$, $ZW$, $Zb\bar{b}$, $Zc\bar{c}$, $Wt\bar{b}$ and $t\bar{t}$. For the SUSY backgrounds all pair production processes involving squarks, gluinos, sleptons, charginos and neutralinos have been included.\\
Two pairs of isolated leptons with opposite sign and the same flavour, with $p_T>$ larger than 10 GeV and within $|\eta|<$ 2.4 are required. The isolation criterion demands no charged particles in with $p_T>$ 1.5 GeV in a cone of $\Delta R<$ 0.3 rad around each lepton track. The invariant mass of the dilepton systems of opposite sign and same flavour are not to be found in an interval of 10 GeV around the $Z$ mass. This $Z$ veto eliminates all backgrounds containing a $Z$ boson. An upper bound of 80 GeV on the $p_T$ of the hardest lepton, an upper limit of 130 GeV on the missing $E_T$ in the event and an upper limit on the $E_T$ of the hardest jet in the event of 100 GeV are efficient against SUSY backgrounds. To reduce the $ZZ$ background further a minimal missing energy of 20 GeV is required. The four lepton invariant mass should not exceed m$_A$-2m$_{\chi^0_1}$.\\
Figure~\ref{fig:disc}(b) shows the 5 $\sigma$ discovery range in the m$_A$-$\tan\beta$ plane for 30 and 100 fb$^{-1}$. The discovery region starts where $\chi^0_2\chi^0_2$ decays become kinematically possible, so for m$_A \geq 2m_{\chi^0_2}$ (here about 230 GeV). The upper reach in m$_A$ is determined by the $A,H$ production cross section. The reach in $\tan\beta$ is determined by the BR of $A,H \to \chi^0_2\chi^0_2$ and $\chi^0_2 \to l^+l^- \chi^0_1$. At 30 fb$^{-1}$ the discovery region reaches m$_A$ about 350 GeV and $\tan\beta$ about 20. For 100 fb$^{-1}$ $\tan\beta$ of 40 can be reached and $A$ masses up to 450 GeV are accessible. The discovery potential is dependent on the choice of the MSSM parameters. We assumed that squarks and gluinos are at the TeV scale. In case where light squark and gluino production is possible there will be extra backgrounds from these channels. It was investigated that this background can be efficiently reduced by some extra cuts on the jet multiplicity ($N_{jet}\leq 2$) and lowering the cut in the transverse energy of the hardest jet in the event to 50 GeV. The branching ratio $A,H \to 4l$ depends on the parameters M$_1$, M$_2$, $\mu$ and m$_{\tilde{l}}$. Large values of $\mu$ and low values of m$_{\tilde{l}}$ seem to be favourable and enhance the decay rate of neutralinos into leptons. If M$_1$ and M$_2$ are small the discovery region in m$_A$ is extended(see Figure~\ref{fig:disc}(b)). We assumed that M$_2$ - M$_1<$ m$_Z$, otherwise the $ZZ$ background rises as the $Z$ veto cannot be used.   
\section{Conclusion}
The fully hadronic $\tau\tau$ channel extends to high Higgs masses. The Higgs mass resolution is the best in this channel. To improve the low Higgs mass range (m$_A <$ 300 GeV) a $\tau$ tagging, based on impact parameters, could be used. Exploiting the $b\bar{b}$ associated production channels with $b$-tagging opens other possibilities and improves significantly the signal to background ratio. The discovery range  of the channel $A,H \to \tau\tau \to l + h + E_T^{miss}$ reaches somewhat lower $\tan\beta$ values. For the boundary values the mass peak is not visible on top of the background strongly dominated by the $Z/\gamma^* \to \tau\tau$ process, which can be sufficiently reduced by $b$ tagging at the cost of the signal. Including the three prong decays in the analyses will enhance statistics.  The neutralino neutralino channel allowing to explore a complementary region in $\tan\beta$-m$_A$ space shows to be very promising.  
\section*{Acknowledgments}
This work was supported through a European Community Marie Curie Fellowship. The European Commission is not responsible for any views or results expressed. The author acknowledges D. Denegri, R. Kinunnen, S. Lehti, F. Moortgat and A. Nikitenko for their inputs, advice and comments which were of great value.

\end{document}